\documentclass[aps,prl,twocolumn,superscriptaddress]{revtex4-2}

\usepackage{amsmath,amssymb}
\usepackage{graphicx}
\usepackage{bm}

\begin{document}

\title{Criticality without Temperature in an Ising Spin System}

\author{C. Kakalis}
\affiliation{Department of Physics, University of Athens, GR-15784, Greece}
\author{F.~K. Diakonos}
\affiliation{Department of Physics, University of Athens, GR-15784, Greece}

\date{\today}

\begin{abstract}
Criticality in the Ising model is conventionally generated by Hamiltonian dynamics and controlled by temperature. Here we show that critical-like behavior can emerge in an Ising system whose dynamics is completely independent of the Ising Hamiltonian and contains no temperature-like parameter. We introduce an adaptive cluster dynamics in which spin connectivity is controlled by a local quiet-time variable—the time elapsed since a spin was last updated. The Ising Hamiltonian enters only as an observable characterizing the resulting configurations. As the parameter $\alpha$, controlling the connectivity strength, is varied, the system undergoes spontaneous symmetry breaking accompanied by strong collective fluctuations. The cluster-size distribution develops an extended power-law regime, terminated by a small finite-size hump at the largest scales, while the Ising energy exhibits a singular response to the variation of $\alpha$. These results show that memory-dependent connectivity alone can generate collective critical behavior, revealing a route to nonequilibrium criticality without energy-based dynamics.
\end{abstract}

\maketitle

\section{Introduction}

The emergence of collective behavior from microscopic dynamics is a central question of statistical physics \cite{Wilson1971, WilsonKogut1974, Goldenfeld1992, Cardy1996}. The Ising model provides a paradigmatic example in which local interactions, together with thermal fluctuations, produce spontaneous symmetry breaking and critical behavior controlled by temperature \cite{Onsager1944}. Collective transitions, however, are not the exclusive domain of equilibrium dynamics. Nonequilibrium systems can generate ordering and critical behavior directly through their dynamical rules, as demonstrated by absorbing-state transitions and driven lattice systems \cite{MarroDickman1999, Hinrichsen2000, Odor2004}. Adaptive networks extend this idea by allowing the connectivity itself to evolve with the dynamical state \cite{GrossBlasius2008, GrossSayama2009}.

Here we consider an Ising spin system whose dynamics is completely decoupled from the Ising Hamiltonian. No temperature, energy-based acceptance criterion, or equilibrium sampling enters the evolution. Instead, spins are organized into clusters through an adaptive connectivity rule controlled by their local quiet time, defined as the time elapsed since their last participation in a cluster update. In particular, the probability of linking two like-oriented spins depends on the quiet times of the participating spins. The dynamics therefore couples the instantaneous connectivity of the system to the temporal history of its individual degrees of freedom.

This construction reverses the conventional role of the Ising Hamiltonian. Rather than generating the dynamics, the Hamiltonian is introduced only a posteriori as an observable characterizing the configurations produced by the adaptive process. Remarkably, varying a single connectivity parameter drives the system from a disordered state to a symmetry-broken regime with pronounced collective fluctuations. The cluster-size distribution evolves from an approximately exponential form into a broad distribution with an extended power-law regime. This regime is followed, at the largest cluster sizes, by a small finite-size hump. Independently, the Ising energy exhibits a singular response to the connectivity strength.

These results demonstrate that critical-like collective behavior can emerge in an Ising spin system without Hamiltonian dynamics or a thermodynamic control parameter. The distinctive ingredient is a memory-dependent connectivity rule in which the probability of cluster formation is regulated by the quiet time of the participating spins. Temporal persistence of individual spins thus provides a dynamical feedback mechanism capable of organizing the system on macroscopic scales and generating collective critical behavior without an underlying energetic dynamics.

\section{Ising spin dynamics adapted to quiet-time}

We consider Ising spins, $s_i=\pm 1$, on a square lattice of size $N\times N$ using an alternative way to control the flipping of individual Ising spins and their arrangement into clusters. 

\subsection{The quiet-time cluster model (QTCM)}

Clusters are constructed similarly to cluster algorithms \cite{SwendsenWang1987, Wolff1989} but the connectivity probability is not determined by a temperature-dependent bond probability. A nearest-neighbor spin $s_j$ of a cluster spin $s_i$ can be added to the cluster only if $s_j=s_i$. The probability of adding this spin to the cluster created at time $t$ (measured in algorithmic steps) is
\begin{equation}
P_{con}(j,t)=1-\exp\left(-\alpha\frac{q(j,t)}{\langle q(t)\rangle}\right),
\label{eq:eq1}
\end{equation}
where $q(j,t)$ denotes the variable {\it quiet time} associated with spin $j$ at time $t$ and $\langle q(t) \rangle=\frac{1}{N^2} \displaystyle{\sum_{i=1}^{N^2}} q(i,t)$ is the average quiet time of all spins before the construction of the cluster. The parameter $\alpha$ controls the connectivity strength. The quiet time of the $i$-th spin evolves in time $t$ according to the rule
\begin{equation}
q(i,t+1)=
\begin{cases}
0,& i \in \mathcal{C}(t) \\
q(i,t)+1,& i \notin \mathcal{C}(t).
\end{cases}
\label{eq:eq2}
\end{equation}
where $\mathcal{C}(t)$ is the cluster created at time $t$. Thus, the system has internal memory: spins that remained inactive for a long time become increasingly likely to participate in a future cluster update. After cluster construction, all spins belonging to the cluster are flipped without any energetic acceptance criterion.

\subsection{Observables}

To phenomenologically characterize the spin dynamics and the generated spin configurations we use global observables which are linear and quadratic functions of the spin variables. In particular, we use (dimensionless) magnetization
\begin{equation}
m=\frac{1}{N^2}\sum_i s_i .
\label{eq:eq3}
\end{equation}
as the linear collective observable and the conventional Ising model energy (also dimensionless here)
\begin{equation}
E=-\sum_{\langle ij\rangle}s_i s_j ,
\label{eq:eq4}
\end{equation}
with $\langle ij \rangle$ indicating the sum over nearest neighbors, as quadratic collective observable.

We also calculate the cluster-size distribution $n(s)$ and the average connectivity probability
\begin{equation}
\overline P(t)=
\frac{1}{N^2}\sum_i P_{con}(i,t) .
\label{eq:eq5}
\end{equation}
We perform simulations for different lattice sizes ($32 \times 32$, $64 \times 64$, $128 \times 128$ and $256 \times 256$) and for different number of algorithmic steps (from $10^6$ to $4 \cdot 10^6$) to check convergence of the obtained numerical results. The control parameter $\alpha$ is varied in the range $[0.5, 1.2]$ which turns out to be sufficient to explore the main phenomenological characteristics of the proposed model.

\section{Simulation Results}

Both dynamical and statistical behavior of the collective observables of the system changes dramatically when the connectivity parameter $\alpha$ is varied. 

\subsection{Magnetization and spontaneous symmetry breaking}

\begin{figure}
\hspace*{-1cm}\includegraphics[width=1.2\linewidth]{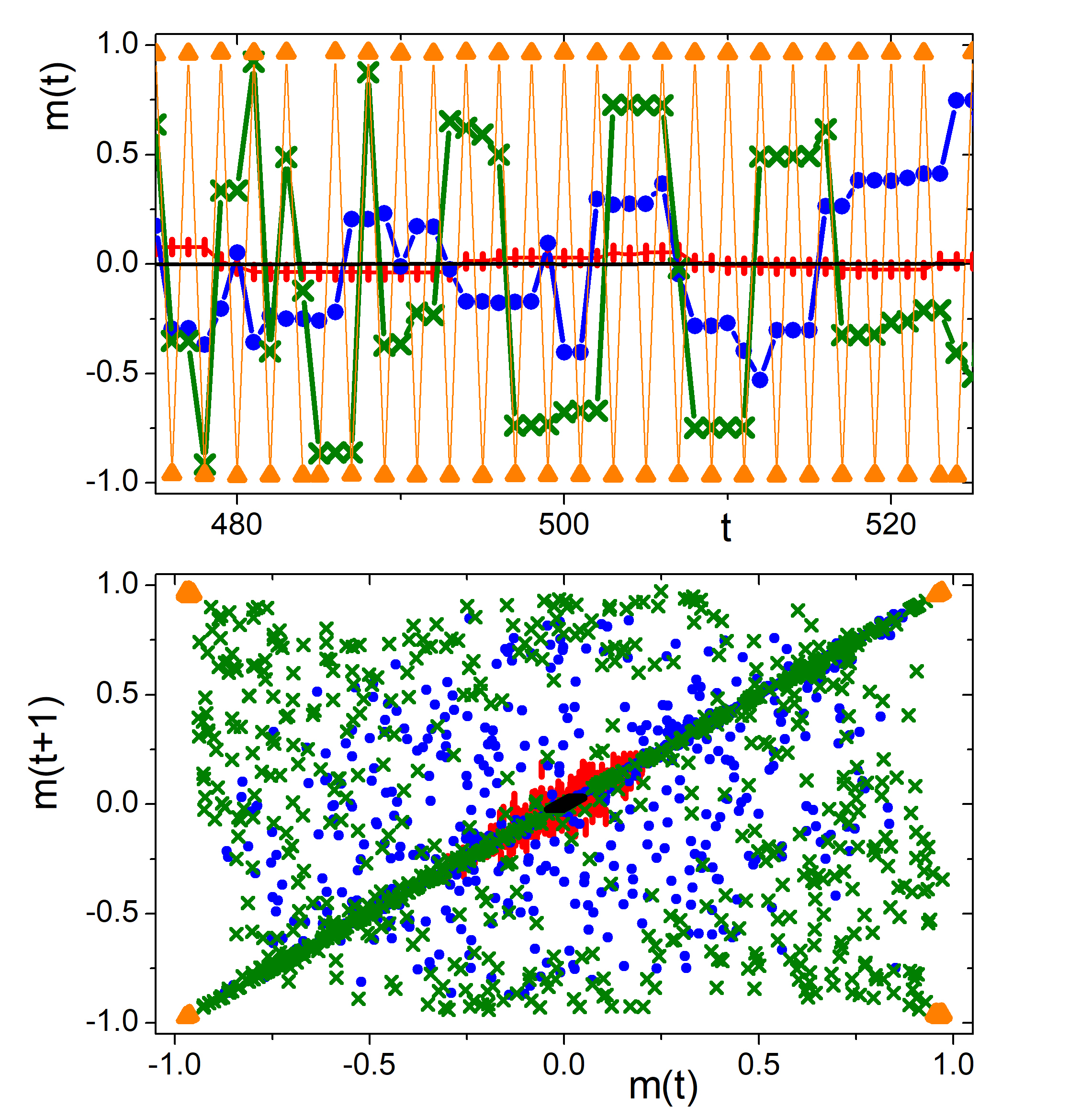}
\caption{Magnetization dynamics in QTCM. In the upper plot is shown a short time interval of 50 algorithmic steps in the time series $m(t)$ for 5 different $\alpha$-values: $\alpha=0.5$ (black dashes, black line), $\alpha=0.8$ (red slabs, red line), $\alpha=1.0$ (blue circles, blue line), $\alpha=1.12$ (olive crosses, olive line) and $\alpha=1.2$ (orange triangles, orange line). In the lower plot the return map $m(t+1)=f(m(t))$ is graphically displayed for the same choice of $\alpha$-values as well as color and symbol code as in the upper plot.}
\label{fig:fig1}
\end{figure}

For small values of $\alpha$ the magnetization fluctuates around zero with a tiny amplitude. As $\alpha$ increases the fluctuation frequency and the corresponding amplitude increases as well. Fluctuations become of order one when 
$\alpha \approx 1.0$ and reach the limiting values $m \approx \pm 1$ when $\alpha \approx 1.12$. When $\alpha \approx 1.2$ the magnetization practically jumps continuously from $m \approx -1$ to $m \approx +1$ and vice versa at high frequency. This behaviour is demonstrated in the upper plot in Fig.~(\ref{fig:fig1}) where a short interval of $50$ algorithmic steps of the magnetization time series for $\alpha=$ $0.5$ (black line), $0.8$ (red slabs, red line), $1.0$ (blue circles, blue line), $1.12$ (olive crosses, olive line)  and $1.2$ (orange triangles, orange line) is shown. In the lower plot of the same figure the return map $m(t+1)=f(m(t))$ in each of these cases of $\alpha$ values is also plotted. In the return map is clearly seen that the underlying magnetization dynamics generated by the introduced model is strongly correlated for $\alpha=0.5$. The magnetization time series is localized around the $(0,0)$ region of the diagonal in $(m(t),m(t+1))$ plane (black dashes, hard to see). For $\alpha=0.8$ the time series covers an elongated region along the diagonal while at the same time the dynamics become more stochastic covering also a region of $(m(t),m(t+1))$-plane around the diagonal (red slabs). The stochastic character of the magnetization dynamics increases drastically as $\alpha$ takes values within the region $[1,1.12]$ (blue circles for $\alpha=1.0$ and olive crosses for $\alpha=1.12$). The magnetization time series tends to cover large fractions of $(m(t),m(t+1))$-plane and for $\alpha \approx 1.12$ it covers almost the entire plane (olive crosses). However, it is worth to notice that a substantial part of the magnetization time series still remains localized along the diagonal $m(t+1)=m(t)$. For $\alpha=1.2$ (orange triangles) the magnetization time series becomes almost periodic. As a result in the return map the trajectory is located around the four edges of $(m(t),m(t+1))$-plane: $(-1,-1)$, $(-1,1)$, $(1,-1)$ and $(1,1)$. 

The associated magnetization probability distributions $\rho(m)$ also reflect this dynamics in a transparent way. For $\alpha=0.5$ the distribution $\rho(m)$ is approximately a narrow Gaussian similar to a delta-spike. Increasing $\alpha$ the magnetization distribution broadens. Around $\alpha\simeq 1$ the distribution becomes similar to that observed at a critical point of a ferromagnetic system at thermal equilibrium, characterized by a higher order maximum at zero. For higher values of $\alpha$, two symmetric peaks develop and move continuously towards $m=\pm 1$, indicating spontaneous symmetry breaking. However, this spin system does not choose a specific orientation since it continuously jumps from negative to positive values and vise versa. When $\alpha$ reaches $\alpha \approx 1.12$, beyond the two dominant peaks, which approach the values $m \pm 1$, two secondary peaks close to the dominant ones also emerge. This behavior is clearly illustrated in Fig.~2 where the magnetization distribution is plotted for five different values of $\alpha$: $0.5$, $0.8$, $1.0$, $1.12$ and $1.2$.
\begin{figure}
\includegraphics[width=\linewidth]{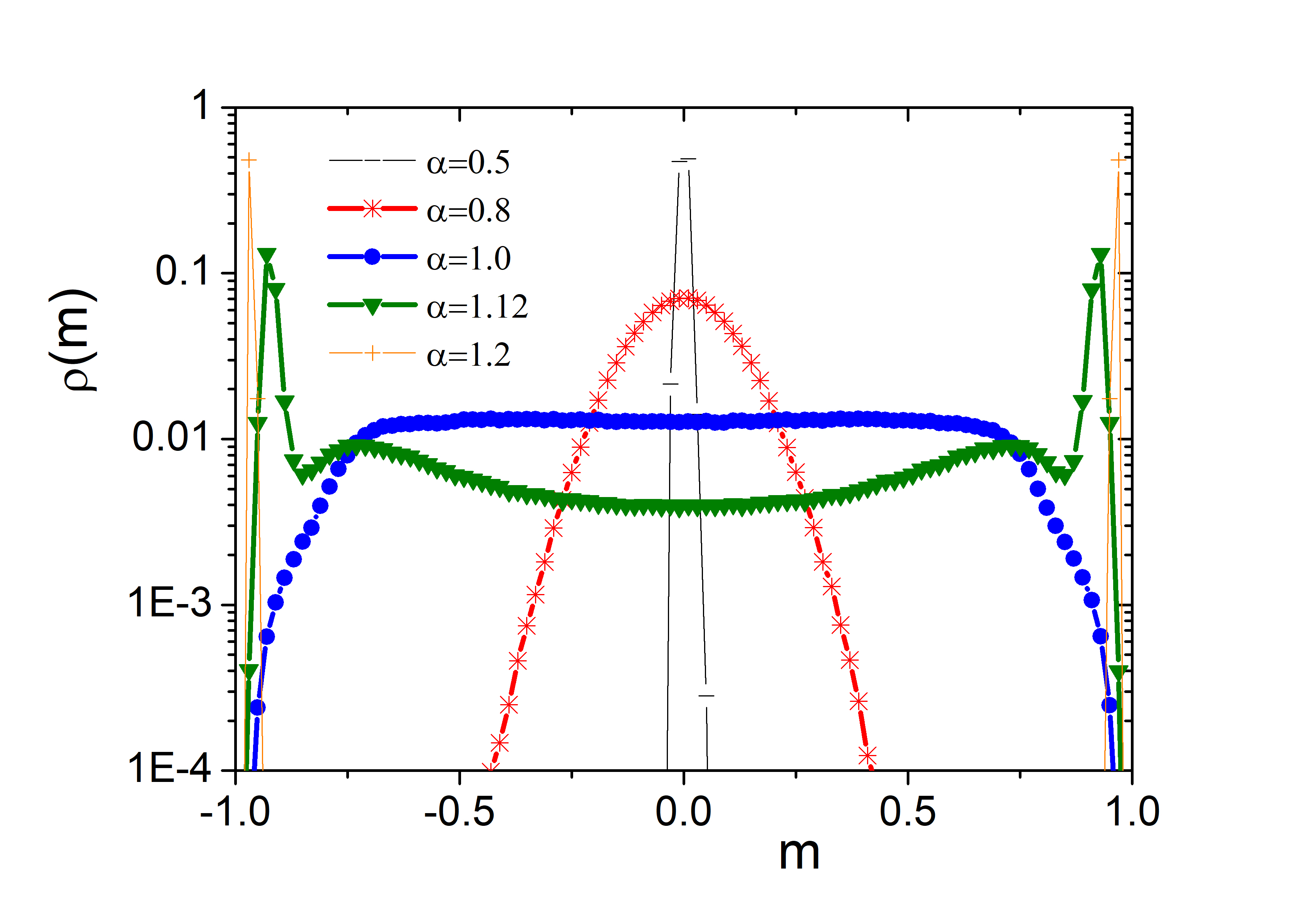}
\caption{Magnetization distributions of the QTCM for different values of $\alpha$ in logarithmic scale. For $\alpha=0.5$ magnetization possesses tiny fluctuations around zero and the corresponding distribution is delta-like concentrated around $m=0$ (black dashes, black solid line). For $a=0.8$ the distribution broadens becoming a Gaussian around $m=0$ with a variance approximately equal to $0.3$ (red stars, red solid line). At $a=1.0$ the  magnetization distribution flattens around $m=0$, indicating the emergence of a higher order maximum (blue circles, blue solid line). When $a=1.12$ the distribution has obtained two dominant maxima symmetric with respect to $m=0$ for values of $m$ close to $\pm1$. These primary maxima are followed by two secondary ones at the values $m \approx \pm 0.75$ (olive inverse triangles, olive solid line). Finally, for $\alpha=1.2$ two symmetric, strongly localized maxima, even closer to $m=\pm 1$ are formed (orange crosses, orange solid line).}
\label{fig:fig2}
\end{figure}

\subsection{Singularity in mean energy}

\begin{figure}
\includegraphics[width=\linewidth]{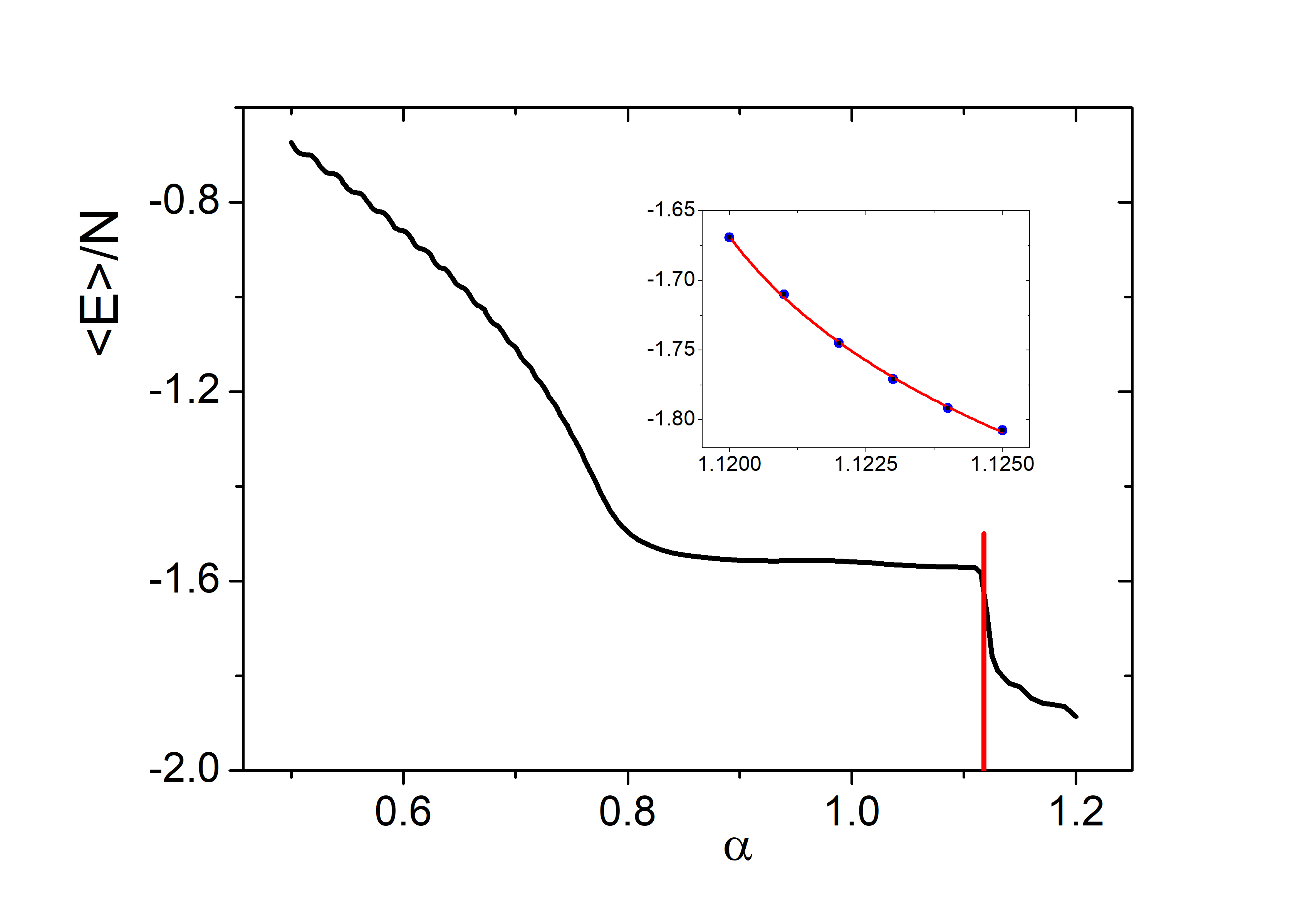}
\caption{Mean energy per spin as a function of $\alpha$ in QTCM. The vertical red line at $\alpha_c \approx 1.11787$ indicates the critical $\alpha$-value. In the inset it is shown a zoom of the main plot in the region just above $\alpha_c$. The blue circles are the calculated points for $\frac{\langle E \rangle}{N}$ while the red line is a power-law fit with exponent $0.067(005)$.}
\label{fig:fig3}
\end{figure}

The energy calculated from Eq.~(\ref{eq:eq4}) exhibits a nonanalytic response with respect to variations of $\alpha$ despite the fact that it is not involved in the dynamics of the spin system. The derivative $-\frac{\partial\langle E\rangle}{\partial\alpha}$
shows a one-sided power-law divergence close to 
$\alpha_s\simeq1.12$ of the form
\begin{equation}
-\frac{\partial\langle E\rangle}{\partial\alpha}
\sim
(\alpha-\alpha_s)^{-0.93}.
\label{eq:eq6}
\end{equation}
This behavior is clearly illustrated in Fig.~(\ref{fig:fig3}) where the dependence of $\langle E \rangle$ on $\alpha$ is shown. The inset contains a zoom in the region around $\alpha_c \approx 1.12$ where the power-law behavior of $\langle E \rangle$ around $\alpha_c$ is observed.

\subsection{Cluster-size distribution}

\begin{figure}
\includegraphics[width=\linewidth]{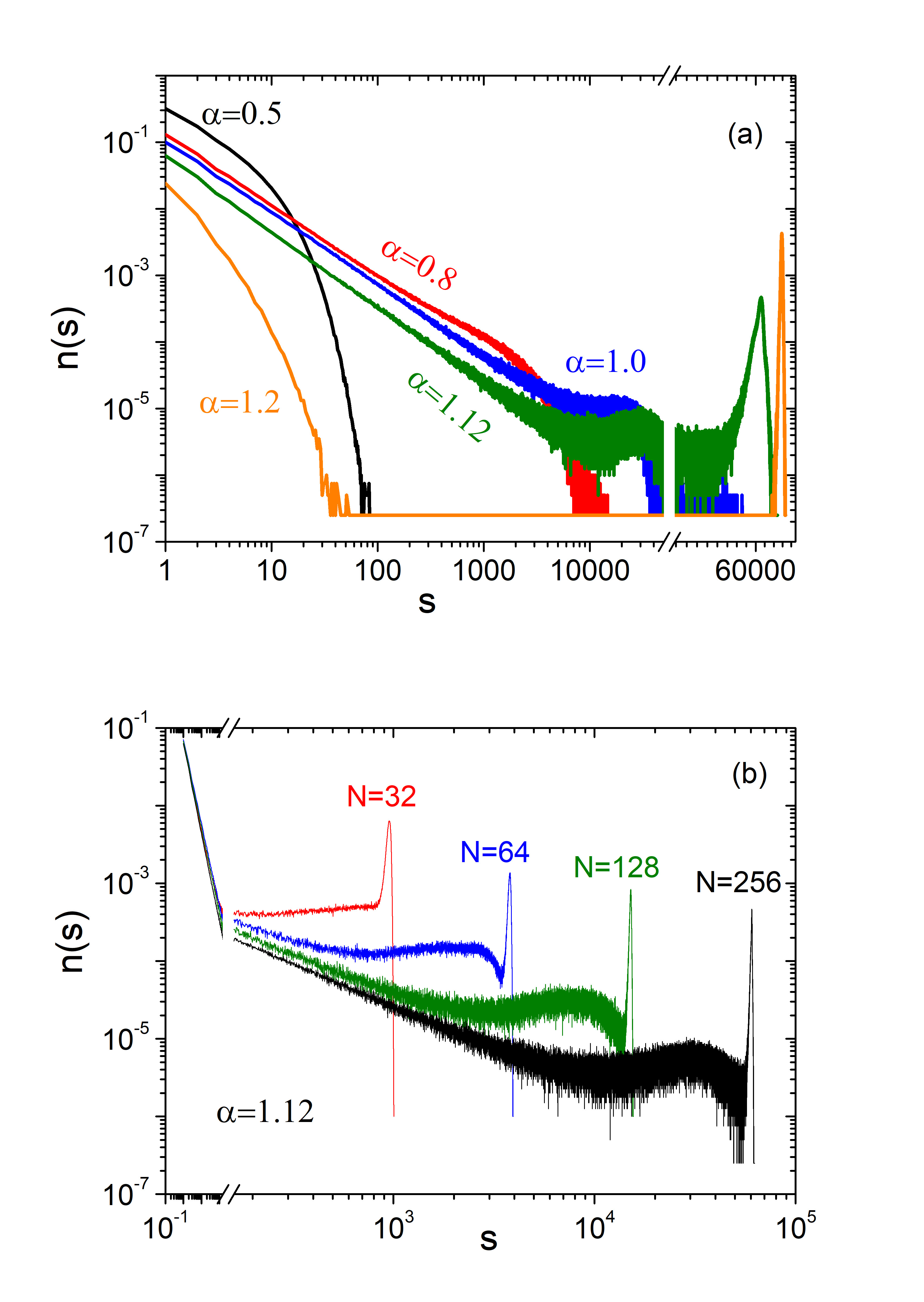}
\caption{(a) The distribution of cluster sizes for the QTCM model using $256^2$ Ising spins for different values of the control parameter $\alpha$: $0.5$ (black line), $0.8$ (red line), $1.0$ (blue line), $1.12$ (olive line) and $1.2$ (orange line). The break on the horizontal axis is at $s=50000$ and it is used to make the plateau region and the hump at large $s$-values more visible. In (b) we show the cluster size distribution for $\alpha=1.12 \approx \alpha_c$) calculated using different number of Ising spins: $32 \times 32$ (black line), $64 \times 64$ (red line), $128 \times 128$ (blue line) and $256 \times 256$ (olive line). The break on the $s$-axis is at $s=160$ and it is used to make the details of the distribution in the large $s$ region more visible.}
\label{fig:fig4}
\end{figure}

The cluster-size distribution provides further evidence for a collective transition. For small $\alpha$ the distribution exhibits an exponential decay $n(s)\sim e^{-s/s_0}$ with $s_0$ being a few decades. As $\alpha$ increases, a power-law regime 
\begin{equation}
n(s)\sim s^{-\tau},
\label{eq:eq7}
\end{equation}
followed by a plateau region ending with an abruptly terminating hump, emerges. Approaching the critical region $[1,1.12]$, the domain of validity of the power-law behavior increases, extending over several decades in cluster size. Above $\alpha_c \approx 1.12$ the cluster-size distribution attains a disconnected profile. In the region containing cluster sizes up to a few hundreds an exponential decay is observed followed by a practically empty region which ends at very high cluster sizes with a narrow and very abrupt hump. This behavior is illustrated in Fig.~(\ref{fig:fig4}a) where the cluster-size distribution for $256^2$ spins and for $\alpha$-values: $0.5$ (black line), $0.8$ (red line), $1.0$ (blue line), $1.12$ (olive line) and $1.2$ (orange line), is shown. Furthermore, at the critical value $\alpha_c \approx 1.12$ it is observed that for spin systems of sizes increasing from $N=20$ to $N=256$, the power-law behavior becomes progressively clearer and the scaling region increases proportional to $N^2$ (see Fig.~(\ref{fig:fig4}b)). Thus, in this case, the hump is clearly a finite-size effect. Notice that in the interval 
$0.9 \lesssim \alpha \lesssim 1.12$ the power-law exponent remains approximately constant, $\tau\simeq -1.08~ \text{to}~-1.11$.

\subsection{Connectivity probability and link to percolation}

\begin{figure}
\includegraphics[width=\linewidth]{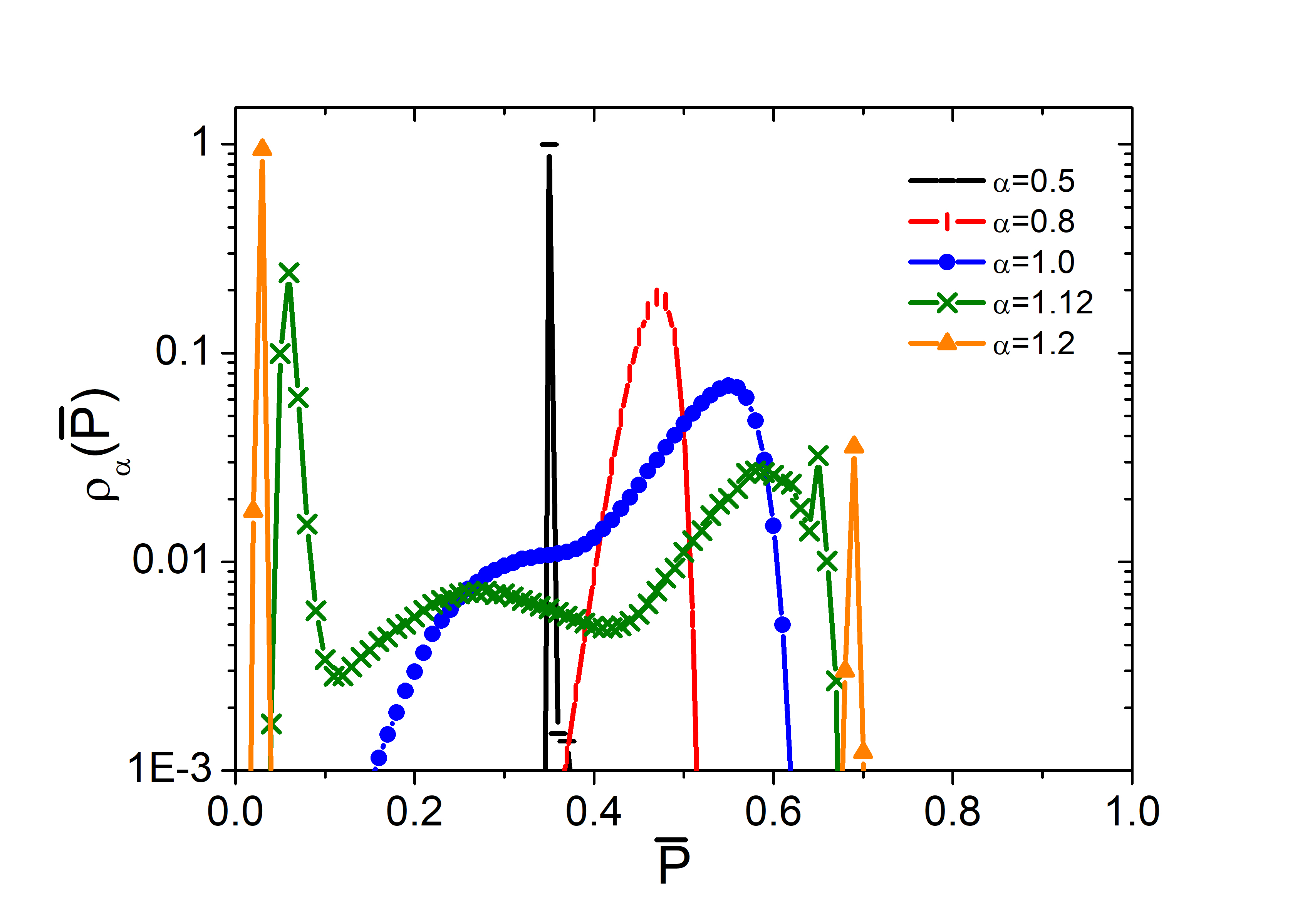}
\caption{The distribution $\rho_{\alpha}(\overline{P})$ of mean connectivity probability values $\overline{P}$ for five different values of $\alpha$: $0.5$ (black dashes, black solid line), $0.8$ (red slabs, red solid line), $1.0$ (blue circles, blue solid line), $1.12$ (olive diagonal crosses, olive line) and $1.2$ (orange triangles, orange solid line).}
\label{fig:fig5}
\end{figure}

Finally, we calculate the mean connectivity probability, for different values of the control parameter $\alpha$, as a function of algorithmic time. From each of these time series we extract the corresponding density $\rho_{\alpha}(\overline{P})$ testing its convergence as the entire time interval is changed from $10^6$ to $4 \cdot 10^6$ algorithmic steps. In Fig.~(\ref{fig:fig5}) we show the results for $\alpha=$ $0.5$, $0.8$, $1.0$, $1.12$ and $1.2$. We observe that $\overline{}{P}$ is practically constant for $\alpha=0.5$ (black dashes, black solid line) taking values close to $0.35$. As $\alpha$ increases the distribution of $\overline{P}$-values broadens and for the case $\alpha=0.8$ the distribution tail includes the value $0.5$ which is the critical bond percolation threshold for a square lattice (red slabs, red solid line) \cite{Kesten1980}. This could explain the appearance of a power-law region in cluster-size distribution for $\alpha=0.8$ (see red line in Fig.~(\ref{fig:fig4}a)). Increasing further $\alpha$ and approaching the critical region $[1,1.12]$ the distribution continues to broaden and the dominant peak moves towards the value $0.6$ which is approximately the site percolation threshold for a square lattice \cite{ReynoldsStanleyKlein1980}. At the same time, the statistical weight of smaller values of $\overline{P}$ -- which are subcritical in 2D-percolation -- increases too. This is clearly seen in Fig.~(\ref{fig:fig5}) through the result obtained for $\alpha=1.0$ (blue circles, blue solid line). For $\alpha=1.12 \approx \alpha_c$ the distribution of values $\overline{P}$ has a central broad peak around the site percolation threshold $0.59$ for 2D-percolation on a square lattice. At the same time, additional peaks develop for lower and higher values of $\overline{P}$. These secondary peaks may be higher than the central one, but their width is smaller (see olive diagonal crosses, olive line in Fig.~(\ref{fig:fig5})). Increasing further $\alpha$ beyond the critical region $[1,1.12]$, the statistical weight of $\overline{P}$-values in the region where 2D-percolation critical values are located ($[0.5,0.6]$) becomes zero and this is in accordance with the fact that in Fig.~(\ref{fig:fig4}a) the power-law region in the cluster-size distribution disappears. In Fig.~(\ref{fig:fig5}) this is clearly seen for $\alpha=1.2$ (orange triangles, orange line) where the distribution of $\overline{P}$-values is shown to posses two narrow maxima at values well below and well beyond the critical domain $[0.5,0.6]$ of 2D-percolation in a square lattice.

\section{Discussion}

The transition reported here is not an equilibrium Ising transition. The Ising Hamiltonian plays no role in the dynamics; instead, its associated observables provide a powerful diagnostic of critical behavior emerging from an entirely different mechanism. The transition is driven by the self-organized evolution of the quiet-time field, which dynamically controls connectivity and promotes the formation of correlated clusters. {\em Critical behavior thus emerges without Hamiltonian-driven dynamics or equilibrium sampling.}

The evolution of the mean connectivity probability further points to a possible connection with percolation. As $\alpha$ increases, $\overline P$ evolves from below the bond-percolation threshold toward values close to the critical probabilities for both bond and site percolation on the square lattice. This proximity suggests a percolation-related mechanism underlying the observed transition, although it does not, by itself, establish universality with either percolation class \cite{StaufferAharony1994, Grimmett1999}. A detailed finite-size scaling analysis and the determination of critical exponents are left for future work.

\section{Conclusion}

We have introduced an adaptive Ising spin dynamics based on quiet-times of the individual spins. These dynamics produce critical behavior in an Ising system without temperature, thermal fluctuations, or Hamiltonian-driven evolution. The emergence of spontaneous symmetry breaking, singular energy response, and scale-free cluster distributions demonstrates that collective critical phenomena can arise solely from
memory-controlled connectivity. 

More broadly, our results challenge the presumption that critical behavior in complex systems (like biological ones) must be rooted in an effective energy landscape, suggesting instead that adaptive temporal memory alone can generate collective criticality.

\end{document}